\documentclass[twocolumn,secnumarabic,nofootinbib,tightenlines,showpacs,floatfix,superscriptaddress,aps,prx]{revtex4-1}
\usepackage{amsfonts,amsbsy,amssymb,amsmath,bm}
\usepackage{color}
\usepackage{txfonts} %otvr needed for not identical
\usepackage[colorlinks=true,linkcolor=blue,filecolor=blue,menucolor=yellow,urlcolor=blue,citecolor=blue,anchorcolor=blue]{hyperref}
\usepackage{graphicx}% Include figure files
\usepackage{dcolumn}% Align table columns on decimal point
\usepackage{times}

\begin{document}
\title{Dissipation-Induced Super Scattering and Lasing $\cal PT$-Spaser}
\author{Simin Feng}
%\email[Corresponding author:\ \ ] {simin.feng@navy.mil}
\email{simin.feng@navy.mil}
\affiliation{Sensor Technology, Naval Surface Warfare Center, Dahlgren, Virginia 22448}
\date{\today}

\begin{abstract}
Giant transmission and reflection of a finite bandwidth are shown to occur at the same wavelength when the electromagnetic wave is incident on a periodic array of $\cal PT$-symmetric dimers embedded in a metallic film.  Remarkably, we found that this phenomenon vanishes if the metallic substrate is lossless while keeping other parameters unchanged.  When the metafilm is adjusted to the vicinity of a spectral singularity, tuning substrate dissipation to a critical value can lead to supper scattering in stark contrast to what would be expected in conventional systems.  The $\cal PT$-synthetic plasmonic metafilm acts as a lasing $\cal PT$-spaser, a planar source of coherent radiation.  The metallic  dissipation provides a mean to couple light out of the dark modes of the $\cal PT$-spaser.  Above a critical gain-loss coupling, the metafilm behaves as a meta-gain medium with the meta-gain atoms made from the $\cal PT$-plasmonic dimers.  This phenomenon implies that super radiation is possible from a cavity having gain elements by tuning the cavity dissipation to a critical value.
\end{abstract}

  %78.20.Ci
\maketitle

\section{Introduction}
In a pioneering work, Bender and colleagues proved that non-Hermitian Hamiltonian with parity-time ($\cal PT$) symmetry may exhibit entirely real spectrum below a phase transition (symmetry breaking) point\cite{Bender1,Bender2}.  Inspired by this emerging concept, in the past decade there has been a growing interest in studying $\cal PT$-symmetric Hamiltonian in the framework of optics\cite{Bendix1}-\cite{Castaldi} where the $\cal PT$ complex potential in quantum mechanics is translated into a complex permittivity profile satisfying $\epsilon(\bm{r})=\epsilon^*(-\bm{r})$ in optical systems.  In optics, most of the $\cal PT$-symmetric structures are realized by parallel waveguides or media with alternating gain and loss either along or across the propagation direction.  The periodic spatial modulation of gain and loss in photonics and plasmonics structures has led to many intriguing phenomena such as nonreciprocal light propagation\cite{Yin,Feng1} and invisibility\cite{Feng2}-\cite{Lin}, power oscillations\cite{Ruter}-\cite{Zheng}, coherent perfect absorptions\cite{Chong1}-\cite{Sun}, loss-induced transparency\cite{Guo}, nonreciprocal Bloch oscillations\cite{Longhi2}, optical switching\cite{Chen}-\cite{Lupu}, unidirectional tunneling\cite{Savoia}, loss-free negative refraction\cite{Fleury}, and laser-absorbers\cite{Longhi3}-\cite{Chong2}.  The $\cal PT$-symmetric systems are a subset of open quantum systems for which Hamiltonian is non-Hermitian and the eigenvalues are complex in general\cite{Rotter1}.  The unique properties associated with non-Hermitian Hamiltonian are exceptional points and spectral singularities.  An exceptional point is a branch point singularity associated with level repulsion\cite{Rotter2}-\cite{Heiss3} and symmetry breaking\cite{Guo,Chong2,Klaiman}-\cite{Larry}.  The existence of the exceptional point has been observed in microwave experiments\cite{Dembow,Bittner}.  Spectral singularity is related to scattering resonance of non-Hermitian Hamiltonian\cite{Mostaf1}-\cite{Mostaf3} and manifest itself as giant transmission and reflection with vanishing bandwidth\cite{Mostaf3,Aalipour}.  Exceptional points and spectral singularities possess interesting electromagnetic properties\cite{Longhi4,Peng} and have attracted much attention lately.  \\

Parity-time synthetic materials represent a new class of metamaterials with novel electromagnetic properties arising from a delicate balance between loss and gain elements.  Global $\cal PT$ symmetry is a demanding condition.  Systems with local $\cal PT$-symmetry are easier to implement.  Array of $\cal PT$-symmetric dimers where each gain-loss pair in itself possesses local $\cal PT$-symmetry with respect to its own center allows for the real spectrum in the right parameter region\cite{Bendix1,Bendix2}.  Except for compensating loss with gain, active plasmonic materials offer an ideal platform for studying non-Hermitian Hamiltonian in the electromagnetic domain at the subwavelength scale.  Most studies on $\cal PT$-symmetric structures use analytical models based on either one dimensional scalar Helmholtz equation or two dimensional scalar paraxial wave equation.  For plasmonic metamaterials having subwavelength ``meta-atom'' as resonators, above analytical descriptions are not applicable.  Nevertheless, the plasmonic metafilms are complex quantum systems with strong coupling and in general can be described by non-Hermitian Hamiltonian.  Based on the spaser concept proposed by Bergman and Stockman\cite{Bergman,Stockman}, Zheludev {\it et. al.}\cite{Zheludev} have suggested ''lasing spaser'' fueled from the dark modes of the spaser via structural symmetry breaking.  Experiments towards this concept have been reported\cite{Plum,Beijnum}.
In this paper, we investigate electromagnetic properties of ${\cal PT}$-synthetic plasmonic metafilm which is composed of a planar array of coupled $\cal PT$-symmetric dimers.  We have found that this structure can display super scattering effect by control of plasmonic substrate dissipation.  When the metafilm is steered to the vicinity of a singular region of the system, tuning the substrate dissipation to a critical value can lead to substantially amplified waves radiated from both sides of the metafilm.  Our device acts as a lasing ${\cal PT}$-spaser.  The giant gain is drawn from the dark modes of the $\cal PT$-spaser via metallic dissipation which breaks the ${\cal PT}$ symmetry and couples light out of the ${\cal PT}$-spaser.  Next, we describe the theoretical approach in section \rm{II} followed by discussion of the super scattering in section \rm{III}, phase transition in section \rm{IV}, physical explanation in section \rm{V}, and summarize our results in the last section.

\section{Theoretical Approach}
\begin{figure}[htb]
\vspace{-2mm}
\includegraphics[width=.46\textwidth]{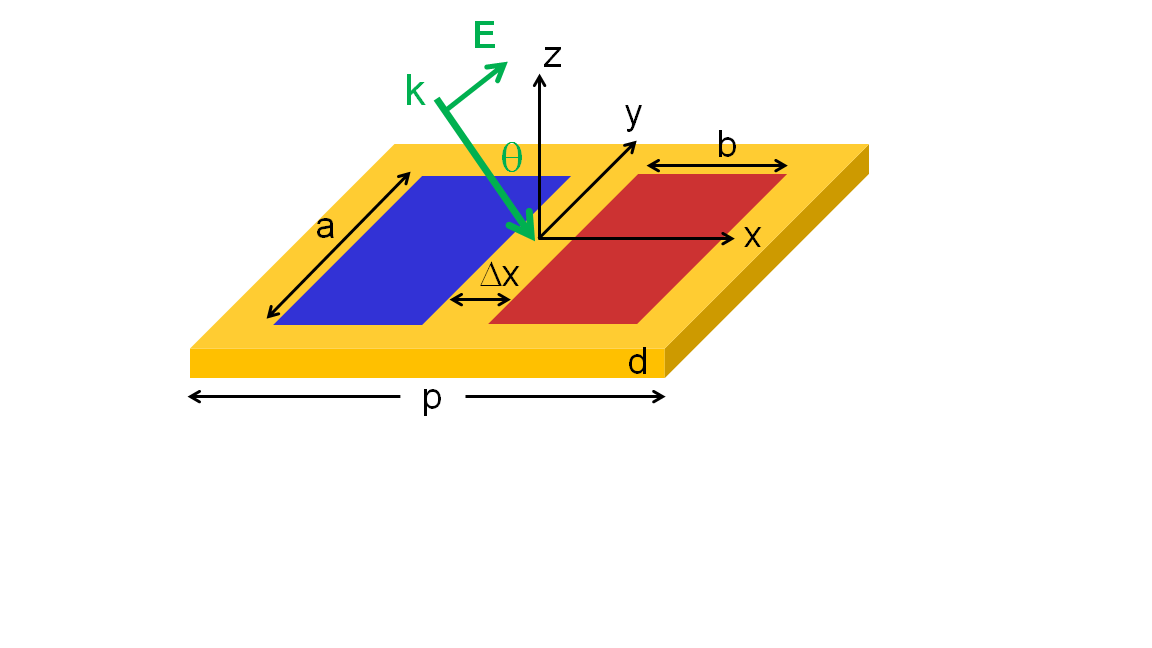}
\vspace{-1.8cm}
\caption{(Color online)  A schematic showing a $\cal PT$-symmetric unit cell composed of loss (blue) and gain (red) subwavelength elements embedded within an aluminum thin-film.  The dimers and the aluminum film have the same thickness, i.e. the metallic mesh is filled with gain-loss elements.  The unit cell repeats in the {\it x-y} plane with the same period in both directions.  The real part of the relative permittivity of the loss and gain elements is fixed at 3.6 through out this work.  The imaginary part varies, but satisfies $\epsilon_{gain}''=-\epsilon_{loss}''$ to ensure the local $\cal PT$ symmetry.  The permeability is unit for all the materials.  The period $p=3.5\,\mu$m, the dimer length $a=2.5\,\mu$m and width $b=1.0\,\mu$m are fixed throughout the paper.  The incidence wave is p-polarized with the electric field parallel to the {\it x-z} plane.}
\label{Fig1Rectwo}
\end{figure}

Figure~\ref{Fig1Rectwo} schematically depicts a unit cell in a planar subwavelength square array of gain-loss elements embedded in a ultra-thin metallic film.  The gain-loss dimer repeats in the {\it x-y} plane.  The plasmonic metafilm satisfies the local $\cal PT$ symmetry with respect to the $x$ directions, i.e. $\epsilon(x,y,z)=\epsilon^*(-x,y,z)$ for $\Delta x/2<|x|<b+\Delta x/2$.  This structure cannot be described by the paraxial wave equation due to the abrupt change of electromagnetic (EM) field at the metal-dielectric interfaces.  We numerically solve Maxwell's equations based on rigorous coupled-wave analysis\cite{Moharam}.  Numerical approaches can handle more complicated structures and guide engineering designs to search for the right parameter combination.  This is important for practical implementation of the extraordinary properties predicted by analytical theory.
%%%%%%%%%%%%%%%%%%%%%%%% Figure 1 %%%%%%%%%%%%%%%%%%%%%%%%%%%%%%

For the infrared $\cal PT$-synthetic materials, the dispersion of metal, which is aluminum (Al) in our case, cannot be neglected.  Assume a harmonic time dependence $\exp(-i\omega t)$ for electromagnetic field, the permittivity of Al was obtained by curve-fitting experimental data\cite{Palik} with a Drude model, 
\begin{equation}
\label{Drude}
\epsilon_m = 1- \frac{\omega_p^2}{\omega^2+i\gamma\omega}, 
\end{equation}
where the ``plasma frequency'' $\omega_p=9.38\ \mu m^{-1}$ and the damping constant $\gamma=0.048\ \mu m^{-1}$.  Maxwell's equations can be recast in a Schr\"{o}dinger-type form:
\begin{equation}
i\frac{\partial}{\partial z} \begin{pmatrix} {\bm E}_t  \\ \bm{\hat z}\times {\bm H}_t \end{pmatrix} = 
	\bm{\widetilde H}\cdot \begin{pmatrix} {\bm E}_t  \\ \bm{\hat z}\times {\bm H}_t \end{pmatrix} \,,
\end{equation}
where the subscript `{\it t}' refers to the transverse $(x,y)$ components of the EM field on the meta-surface; and $\bm{\hat z}$ is the unit vector along the $z$ direction.  The Hamiltonian is given by
\begin{equation}
\label{Hamilt}
\bm{\widetilde H} = 
\begin{pmatrix} {\bm 0} & &  k_0\mu_t \bm{\hat I}_t + \displaystyle
	\frac{1}{k_0}\nabla_t \frac{1}{\epsilon_z}\nabla_t  \\
	k_0\epsilon_t\bm{\hat I}_t + \displaystyle\frac{1}{k_0} \bm{\hat z}\times\nabla_t
	\frac{1}{\mu_z} \bm{\hat z}\times\nabla_t  & & {\bm 0}
\end{pmatrix}  \,,
\end{equation}
where $k_0=\omega/c$, and the $c$ is the speed of light in vacuum.  The subscript `{\it z}' refers to the component in the $z$ direction.  
$\bm{\hat I}_t=\bm{\hat I}-\bm{\hat z}\bm{\hat z}$ is the two-dimensional unit dyadic; and
\begin{equation}
\nabla_t\equiv \bm{\hat x}\frac{\partial}{\partial x} + \bm{\hat y}\frac{\partial}{\partial y} \,.
\end{equation}
The form of Eq.~(\ref{Hamilt}) can handle uniaxial anisotropic materials with the optical axis along the $z$ direction.  The Hamiltonian given by Eq.~(\ref{Hamilt}) is non-Hermitian.  The scattering and transfer matrices, as well as the transmittance and the reflectance are calculated numerically.  The transfer matrix which connects the field at the output and the input surfaces is defined as
\begin{equation}
\label{}
 \vert\Psi_i\rangle = \begin{pmatrix}  M_{11} &  M_{12} \\  M_{21} &  M_{22} \end{pmatrix} \vert\Psi_o\rangle  \,,
\end{equation}
where the subscripts 'o' and 'i' refer to the EM field at the output and input surfaces, respectively;  and
\begin{equation}
\vert\Psi\rangle \equiv \begin{pmatrix} {\bm E}_t  \\ \bm{\hat z}\times {\bm H}_t \end{pmatrix}  \,.
\end{equation}
The relationship between the transfer and scattering matrices in our case is given by
\begin{eqnarray}
\label{Spar}
\begin{split}
S_{11} &= M_{21} M_{11}^{-1}  \,, \hspace{.3in}   S_{21} = M_{11}^{-1}   \,, \cr
S_{22} &= -M_{11}^{-1} M_{12}  \,, \hspace{.2in}  S_{12} = M_{22} - M_{21} M_{11}^{-1} M_{12}  \,,
\end{split}
\end{eqnarray}
where the $S_{11}$ and $S_{21}$ are, respectively, the reflection and transmission coefficients of the electric field.  In general, the transfer and scattering matrices are multidimensional due to multiple scattering channels.  We have conducted extensive numerical studies and confirmed that for our geometry the magnitudes of the higher order and cross-polarization scatterings are much smaller than that of the first order event due to the subwavelength nature of the films.  Therefore, the transfer and scattering matrices can be reduced to $2\times2$ matrices.  The problem can be effectively described in a two-dimensional space.  The eigenvalues of the transfer matrix is given by
\begin{equation}
\label{Teign}
\eta^t_\pm = \frac{M_{11}+M_{22}}{2} \pm \sqrt{\left(\frac{M_{11}+M_{22}}{2}\right)^2-1}  \,.
\end{equation}
Here the identity $\det(\bm{M})=1$ (which has been numerically validated under various conditions) has been used.  From Eq.~(\ref{Teign}), we have $|\eta^t_+\eta^t_-|=1$ (which has been numerically tested for various parameters).  The symmetry of the structure requires the relationship $S_{21}=S_{12}$ and $M_{12}=-M_{21}$.  These two conditions have also been confirmed numerically under various conditions.  By using these conditions, the eigenvalues of the scattering matrix can be derived as
\begin{equation}
\label{Seign}
\eta^s_\pm = \frac{M_{21}\pm1}{M_{11}}  \,.
\end{equation}
The eigenfunctions of the non-Hermitian Hamiltonian satisfy bi-orthogonal relationship.  The right-hand eigenvectors of the transfer matrix are given by
\begin{equation}
|\psi^r_\pm\rangle = \begin{pmatrix}  \displaystyle\frac{\eta^t_\pm-M_{22}}{M_{21}}  \\ \\  1 \end{pmatrix} \,,
\end{equation}
and the left-hand eigenvectors
\begin{equation}
\langle\psi^l_\pm| = \begin{pmatrix}  \displaystyle\frac{\eta^t_\pm-M_{22}}{M_{12}}  & &  1 \end{pmatrix} 
\end{equation}
with the property 
\begin{equation}
\langle\psi^l_+|\psi^r_-\rangle = \langle\psi^l_-|\psi^r_+\rangle = 0  \,.
\end{equation}
Above bi-orthogonality has been numerically verified under various conditions.  The transfer and scattering matrices and their eigenvalues and eigenvectors are useful tools for the analysis of exceptional points and spectral singularities.

\section{Super Scattering}
\begin{figure}[htb]
\includegraphics[width=.37\textwidth]{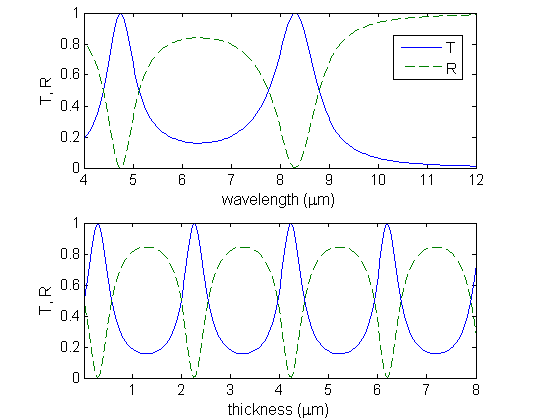}
\caption{(Color online)  Transmittance (blue solid) and reflectance (green dashed) of the normal incidence wave on a lossless metafilm versus wavelength (upper panel) and thickness (lower panel) with the electric field parallel to the shorter edge of the dimers.  The separation of the two dimers $\Delta x=0.5\,\mu$m for both cases.  The thickness of the metafilm $d=1.5\,\mu$m for the upper panel and the wavelength $\lambda=6\,\mu$m for the lower panel.  The relative permittivity of the dimers is real and given by $\epsilon_r=3.6$.}
\label{Fig2GLrec2Tz3}
\end{figure}

In the geometry of Fig.~\ref{Fig1Rectwo}, many parameters can be changed.  In this work, the period ($p=3.5\,\mu m$) of the square array, the size ($a\times b=2.5\times1.0\,\mu m^2$) of the dimers, and the real part of the relative permittivity ($\epsilon'_r=3.6$) of the dimers are fixed throughout the paper.  For a comparison we show in Fig.~\ref{Fig2GLrec2Tz3} the transmittance $(T)$ and reflectance $(R)$ of a normal incident electromagnetic wave onto a lossless metafilm with lossless dimers.  The frequency dependent permittivity of aluminum was taken from the real part of Drude model given by Eq.~(\ref{Drude}).  In the lossless case we have $T+R=1$ (energy conservation) which is clearly demonstrated in Fig.~\ref{Fig2GLrec2Tz3}.  An increase of the transmission is accompanied by a decrease of the reflection and vice versa.  The peaks and valleys of the transmittance and the reflectance repeat periodically with the variation of the thickness of the film.  The metafilm behaves like a low-Q Fabry-P\'{e}rot cavity below $9\,\mu$m.  Above $12\,\mu$m, the metafilm behaves towards a perfect electric conductor (PEC).
%%%%%%%%%%%%%%%%%%%%%%%% Figure 2 %%%%%%%%%%%%%%%%%%%%%%%%%%%%%%
%%%%%%%%%%%%%%%%%%%%%%%% Figure 3 %%%%%%%%%%%%%%%%%%%%%%%%%%%%%%
\begin{figure}[htb]
\includegraphics[width=.35\textwidth]{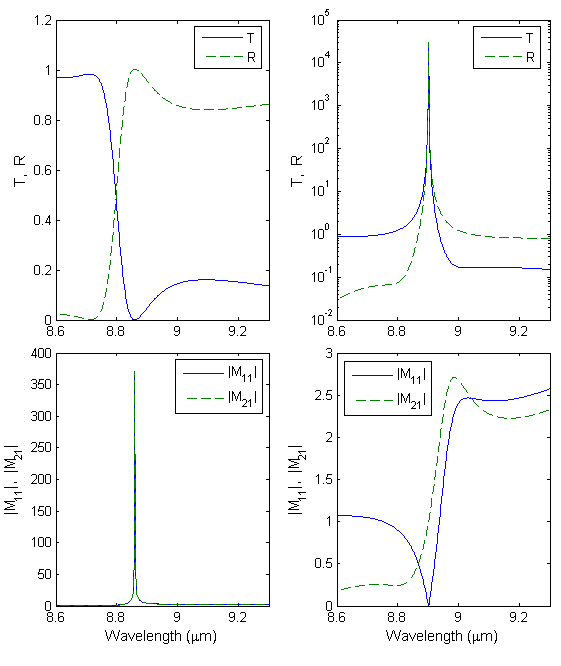}
\caption{(Color online)  Simulation without (left panels) and with (right panels) metallic substrate dissipation.  Upper panels: Transmittance (blue solid) and reflectance (green dashed) for the normal incidence with the electric field parallel to the shorter edge of the dimers.  Lower panels: The magnitude of $M_{11}$ (blue solid) and $M_{21}$ (green dashed).  The separation of the two dimers $\Delta x=0.5\,\mu$m; and the thickness of the mesh $d=1.5\,\mu$m.  The relative permittivity of the gain/loss elements $\epsilon=3.6(1\pm0.06i)$.}
\label{Fig3GLrec2Twss4}
\end{figure}
%%%%%%%%%%%%%%%%%%%%%%%% Figure 4 %%%%%%%%%%%%%%%%%%%%%%%%%%%%%%
\begin{figure}[htb]
\includegraphics[width=.35\textwidth]{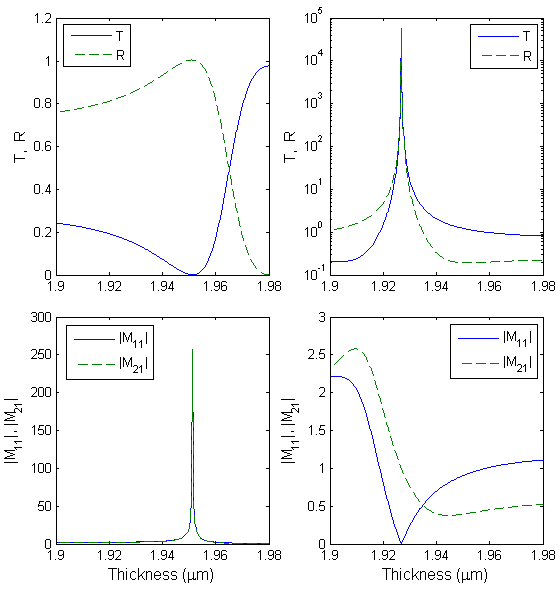}
\caption{(Color online)  Simulation without (left panels) and with (right panels) metallic substrate dissipation.  Upper panels: Transmittance (blue solid) and reflectance (green dashed) vs. the thickness of the metafilm at the wavelength ($\lambda=6\,\mu$m) for the normal incidence with the electric field parallel to the shorter edge of the dimers.  The corresponding analysis of the transfer matrix elements is given in the lower panels.  The separation of the two dimers $\Delta x=0.5\,\mu$m.  The relative permittivity of the dimers $\epsilon=3.6(1\pm0.056i)$.}
\label{Fig4GLrec2Tdss4}
\end{figure}
%%%%%%%%%%%%%%%%%%%%%%%% Figure 5 %%%%%%%%%%%%%%%%%%%%%%%%%%%%%%
\begin{figure}[htb]
\includegraphics[width=.37\textwidth]{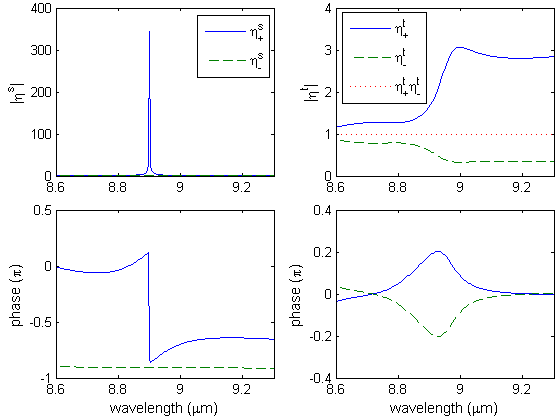}
\caption{(Color online)  Magnitudes (upper panels) and phases (lower panels) of the eigenvalues of the scattering (left panels) and transfer (right panels) matrices corresponding to the condition of Fig.~\ref{Fig3GLrec2Twss4}.  One of the eigenvalues of the scattering matrix diverges with a phase jump (blue solid) at the wavelength $8.92\,\mu$m where the super scattering occurs.  The eigenvalues of the transfer matrix satisfy the condition $|\eta^t_+\eta^t_-|=1$ (red dots) with one magnitude greater than one (blue solid, gain mode) and the other less than one (green dashed, loss mode).}
\label{Fig5GLrec2Twss3}
\end{figure}
%%%%%%%%%%%%%%%%%%%%%%%% Figure 6 %%%%%%%%%%%%%%%%%%%%%%%%%%%%%%
\begin{figure}[htb]
\includegraphics[width=.37\textwidth]{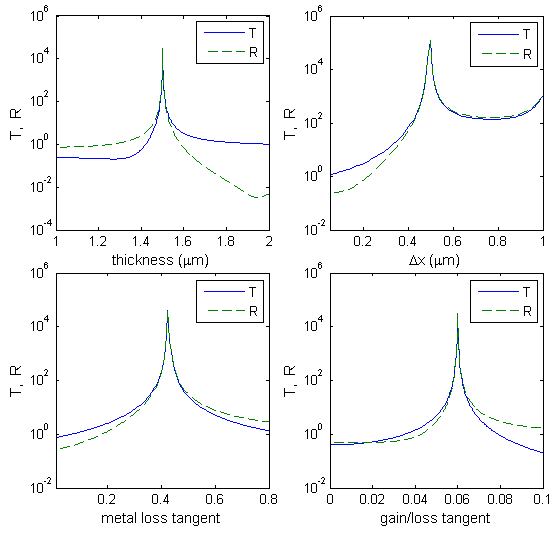}
\caption{(Color online)  Transmittance (blue solid) and reflectance (green dashed) at the wavelength $8.92\,\mu$m versus thickness (upper left), dimer separation (upper right), metal loss tangent (lower left), and dimer loss tangent (lower right).  Other parameters are the same as those in Fig.~\ref{Fig3GLrec2Twss4}.}
\label{Fig6GLrec2Twss5}
\end{figure}
%%%%%%%%%%%%%%%%%%%%%%%% Figure 7 %%%%%%%%%%%%%%%%%%%%%%%%%%%%%%
\begin{figure}[htb]
\includegraphics[width=.37\textwidth]{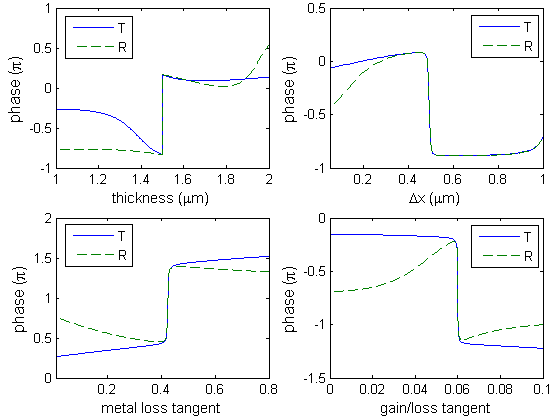}
\caption{(Color online)  Phase variation of the transmitted (blue solid) and reflected (green dashed) electric fields at the wavelength $8.92\,\mu$m versus mesh thickness (upper left), dimer separation (upper right), metal loss tangent (lower left), and dimer loss tangent (lower right).  Each panel is correlated to that in Fig.~\ref{Fig6GLrec2Twss5}.}
\label{Fig7GLrec2Twss6}
\end{figure}
%%%%%%%%%%%%%%%%%%%%%%%% Figure 8 %%%%%%%%%%%%%%%%%%%%%%%%%%%%%%
\begin{figure}[htb]
\includegraphics[width=.33\textwidth]{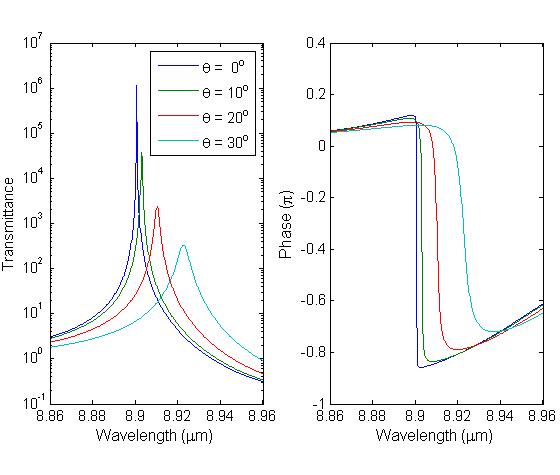}
\caption{(Color online)  Transmittance (left panel) of the p-polarized wave and corresponding phase jump (right panel) of the transmitted electric field for different incidence polar angles from $0^\circ$ to $30^\circ$ at the step of $10^\circ$.  The structure is optimized for the normal incidence with the same parameters as those in Fig.~\ref{Fig3GLrec2Twss4}.}
\label{Fig8GLrec2Tang2}
\end{figure}
%%%%%%%%%%%%%%%%%%%%%%%% Figure 9 %%%%%%%%%%%%%%%%%%%%%%%%%%%%%%
\begin{figure}[htb]
\includegraphics[width=.35\textwidth]{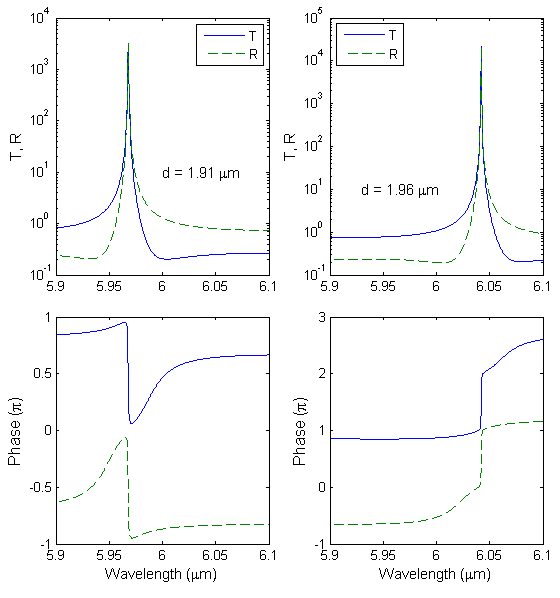}
\caption{(Color online)  Tuning the resonant frequency of the super scattering by varying the thickness of the mesh $d=1.91\,\mu$m (left panels) and $d=1.98\,\mu$m (right panels).  Upper panels: transmittance (blue solid) and reflectance (green dashed) of the normal incident wave.  Lower panels: the corresponding phase jump of transmitted (blue solid) and reflected (green dashed) electric fields.  Other simulation parameters are the same as those in Fig.~\ref{Fig4GLrec2Tdss4}.}
\label{Fig9GLrec2Ttun1}
\end{figure}

Now we proceed to investigate the EM scattering from a plasmonic thin-film having balanced gain-loss elements.  The $\cal PT$-synthetic plasmonic metafilm no longer behaves as a Fabry-P\'{e}rot cavity.  
To demonstrate the effect of the substrate dissipation, we investigate the transmittance and reflectance of the metafilm with and without plasmonic dissipation in the presence of the balanced gain-loss dimers.  The upper left panel in Fig.~\ref{Fig3GLrec2Twss4} shows the transmittance and reflectance of the EM wave normally incident on the $\cal PT$-synthetic lossless plasmonic metafilm of which the dispersion is given by the real part of Drude model in Eq.~(\ref{Drude}).  The relative permittivity of the dimers is given by $\epsilon=3.6(1\pm0.06i)$.  Clearly, the sum of the transmittance and reflectance is close to unit, i.e. $T+R=1$.  Overall the metafilm performs as a conventional lossless medium where the maximum transmission is accompanied by the minimum reflection, and vice versa.  This behavior can be understood as the subwavelength dimers have balanced gain-loss profile and are embedded in the lossless substrate.  The situation changes dramatically when the plasmonic substrate dissipation is taken into account as shown in the upper right panel of Fig.~\ref{Fig3GLrec2Twss4} where the giant transmittance and reflectance occur at the same wavelength, unlike conventional media where the increase of one at the expense of the other.  This peculiar property can be analyzed through the scattering parameters $S_{11}$ and $S_{21}$ that have a common denominator $M_{11}$ as shown by Eq.~(\ref{Spar}).  When the denominator of $S_{11}$ and $S_{21}$ vanishes, both transmittance and reflectance approach infinity as long as $M_{21}$ is finite.  The lower panels in Fig.~\ref{Fig3GLrec2Twss4} show the magnitude of $M_{11}$ and $M_{21}$ with and without the substrate dissipation described by the Drude model in Eq.~(\ref{Drude}).  Without the substrate dissipation (lower left panel), both magnitude of $M_{11}$ and $M_{21}$ are large at the wavelength about $8.88\,\mu$m which explains the null in the transmittance and the peak in the reflectance (see upper left panel).  When turn on the substrate dissipation (lower right panel), the $|M_{11}|$ vanishes at the wavelength about $8.92\,\mu$m whereas the $|M_{21}|$ is finite.  Thus, both transmission and reflection approach infinity.  Here the spectral singularities are manifested as the dissipation-induced super scattering.  
This dissipation-induced super scattering is also observed when varying the thickness of the metafilm at the fixed wavelength $\lambda=6\,\mu$m as illustrated in the upper panels of Fig.~\ref{Fig4GLrec2Tdss4} with the corresponding analysis of the transfer matrix elements given in the lower panels of Fig.~\ref{Fig4GLrec2Tdss4}.  Without the substrate dissipation, the metafilm behaves as a conventional medium where the peak of the transmittance is accompanied by the valley of the reflectance and vice versa (upper left panel).  The production of gain and loss is balanced.  
The plasmonic substrate dissipation induces over-production of gain, leading to super scattering in both forward and backward directions (upper right panel).  As shown in the lower right panel of Fig.~\ref{Fig4GLrec2Tdss4}, the singular behavior is related to the vanishing of the transfer matrix element $M_{11}$ (blue solid), the common denominator of $S_{11}$ and $S_{21}$.  \\

The spectral singularity and the associated giant scattering is one of the signatures of non-Hermitian Hamiltonian of the gain medium.  From Eq.~(\ref{Seign}), when $M_{11}\longrightarrow0$, one eigenvalue diverges and the other one is finite, i.e.
\begin{equation}
\eta^s_+\longrightarrow\infty,  \hspace{.3in}  \eta^s_-\longrightarrow -\frac{M_{22}}{2}  \,.
\end{equation}
This singularity is demonstrated in the upper left panel (blue solid curve) of Fig.~\ref{Fig5GLrec2Twss3} which displays the eigenvalues of the scattering (left panels) and transfer (right panels) matrices with the simulation condition corresponding to that of Fig.~\ref{Fig3GLrec2Twss4}.  There is also a phase jump for the diverged eigenvalue (blue solid curve in the lower left panel).  The eigenvalues of the transfer matrix satisfy the condition $|\eta^t_+\eta^t_-|=1$ (see Eq.~(\ref{Teign})) with one magnitude greater than one and the other less than one, corresponding to the amplification and decay modes, respectively. 
The relationship $|\eta^t_+\eta^t_-|=1$ has been tested for various parameters with and without the substrate dissipation.
Unlike the zero-bandwidth resonance in the longitudinal $\cal PT$-symmetric structures\cite{Mostaf1}-\cite{Aalipour}, the super scattering in our geometry with the transverse local $\cal PT$-symmetry has a finite bandwidth which is easier to demonstrate experimentally and opens up various potential applications.  \\

The coexistence of super transmission and reflection is a manifestation of the resonant state of the complex scattering potential, one of the essential features of non-Hermitian Hamiltonian with gain.  This phenomenon is fundamentally different from the geometric related resonances.  The spectral singularity-induced resonance resides in a localized region of a multidimensional parameter space of the system as implied in Fig.~\ref{Fig6GLrec2Twss5} which shows the super scattering at the wavelength $8.92\,\mu$m versus the geometric parameters and the loss tangents of the substrate and dimers.  At the supper scattering, the transmitted and reflected electric fields experience a $\pi$-phase shift as the parameters across the singularity as shown in Fig.~\ref{Fig7GLrec2Twss6}.  Figure~\ref{Fig8GLrec2Tang2} demonstrates the giant transmittance and the corresponding phase shift of the electric field at different (polar) angles of incidence.  The transmittance decreases with the increase of the incidence angle.  Similar behavior occurs to the reflectance (not show).  Figure~\ref{Fig9GLrec2Ttun1} shows tuning the resonant frequency of the super scattering and the corresponding phase jump of the transmitted and reflected electric fields when varying the thickness of the mesh while keeping other parameters unchanged.

\section{Phase Transition}
\begin{figure}[htb]
\includegraphics[width=.35\textwidth]{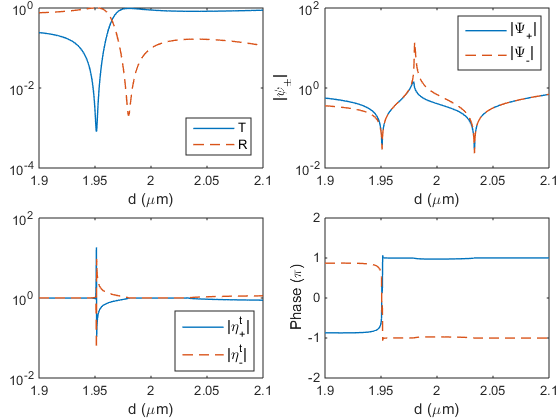}
\caption{(Color online)  Transmittance and reflectance (upper left), norm of eigenstates (upper right), magnitude (lower left) and phase (lower right) of eigenvalues of transfer matrix versus the thickness of the film in the absence of substrate dissipation.  The eigenvalues and eigenfunctions show a series of phase transitions as the thickness changes.  The two dips in the norm of the eigenfunctions indicate the exceptional points while the peak implies a diabolic point.  This three points correlate with the three bifurcation points in the lower left panel.  Wavelength $\lambda=6\,\mu$m, the dimer separation $\Delta x=0.5\,\mu$m.  The relative permittivity of the dimers $\epsilon=3.6(1\pm0.056i)$.}
\label{Fig10GLrec2Tdep1}
\end{figure}

Phase transition is a common feature of open quantum systems.  The plasmonic metafilm is a two-dimensional open quantum system which inherits several types of singularities.  An exceptional point (EP) is characterized by coalescence of both eigenvalues and their associated eigenvectors\cite{Heiss3}.  At the EP, the system reduces to one dimension and the norm of the corresponding eigenfunction vanishes\cite{Heiss3}.  On the contrary, only eigenvalues show degeneracy at the diabolic point.  Without the background dissipation, the metafilm satisfies the ${\cal PT}$ symmetry with balanced gain and loss arranged in the anti-symmetric distribution of the imaginary part of the permittivity.  The two eigenstates of the film are the forward and backward waves.  Their constructive and destructive interferences modify transmission and reflection, as well as phase transitions.
Figure~\ref{Fig10GLrec2Tdep1} shows that the ${\cal PT}$-symmetric metafilm undergoes a multitude of spontaneous symmetry-breaking transitions when varying the thickness.  In the lower left panel of Fig.~\ref{Fig10GLrec2Tdep1}, the magnitudes of the eigenvalues provide the information on the transitions from a norm preservation phase when the thickness $d\in[1.90, 1.95]\,\mu$m to a symmetry-breaking phase with a pairs of eigenstates exhibiting amplification and loss when $d\in[1.95, 1.98]\,\mu$m, and transitions back to the norm conservation phase when $d\in[1.98, 2.03]\,\mu$m, and then back to the broken phase again with net gain and loss in the eigenstates when $d\in[2.03, 2.1]\,\mu$m.  The three bifurcation points in the eigenvalues correlate with the valleys and peak in the norm of the eigenfunctions.  The eigenvalues form an inverse-conjugate pair.  The onsets of the phase transitions arise from the spontaneous ${\cal PT}$-symmetry breaking when one of the system parameters changes.
%%%%%%%%%%%%%%%%%%%%%%%% Figure 10 %%%%%%%%%%%%%%%%%%%%%%%%%%%%%%
%%%%%%%%%%%%%%%%%%%%%%%% Figure 11 %%%%%%%%%%%%%%%%%%%%%%%%%%%%%%
\begin{figure}[htb]
\includegraphics[width=.35\textwidth]{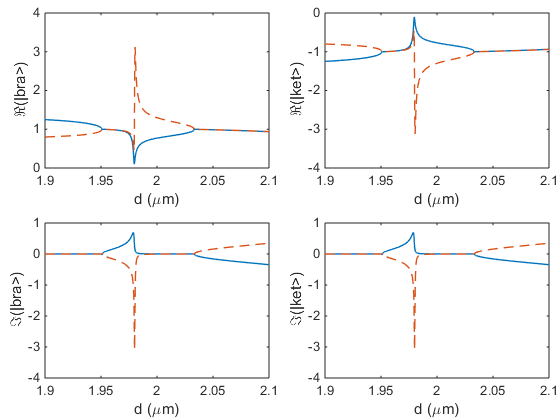}
\caption{(Color online)  Eigenfunctions vs. thickness.  Real (upper panels) and imaginary (lower panels) parts of left (left panels) and right (right panels) hand eigenfunctions.  Showing the two EPs at $d=1.951\,\mu$m and $2.034\,\mu$m and one diabolic point at $d=1.98\,\mu$m.  Same simulation parameters as those in Fig.~\ref{Fig10GLrec2Tdep1}.}
\label{Fig11GLrec2Tdep2}
\end{figure}
%%%%%%%%%%%%%%%%%%%%%%%% Figure 12 %%%%%%%%%%%%%%%%%%%%%%%%%%%%%%
\begin{figure}[htb]
\includegraphics[width=.35\textwidth]{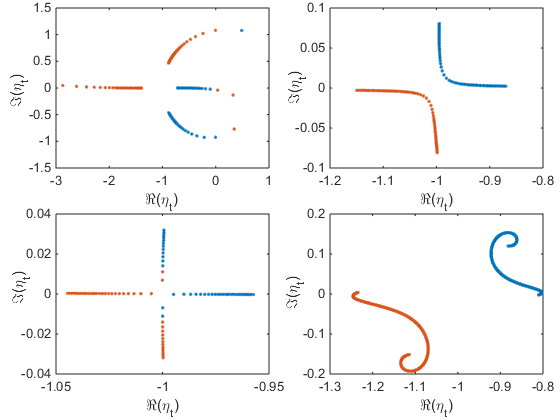}
\caption{(Color online)  Real vs. imaginary parts of eigenvalues of transfer matrix when the thickness varies in the vicinity of the three bifurcation points $d=1.951\,\mu$m (upper left), $d=1.98\,\mu$m (upper right), and $d=2.034\,\mu$m (lower left).  A level repulsion occurs at the $d=1.98\,\mu$m.  The lower right panel shows the real vs. imaginary parts of the eigenvalues when the thickness varying from $1.9\,\mu$m to $2.1\,\mu$m in the presence of the substrate dissipation.  This parameter covers a broader range which includes the ranges in the other three panels.  The blue and red curves represent the loss and gain modes, respectively.  The eigenvalues show less structures in the presence of dissipation compared to those without the dissipation.  Other parameters are the same as those in Fig.~\ref{Fig10GLrec2Tdep1}.}
\label{Fig12GLrec2Tdep3}
\end{figure}
%%%%%%%%%%%%%%%%%%%%%%%% Figure 13 %%%%%%%%%%%%%%%%%%%%%%%%%%%%%%
\begin{figure}[htb]
\includegraphics[width=.38\textwidth]{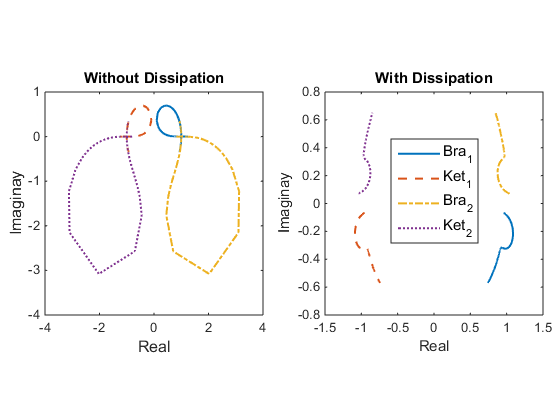}
\caption{(Color online)  Imaginary vs. real parts of eigenfunctions when the thickness varies from $1.9\,\mu$m to $2.1\,\mu$m with and without the metallic dissipation.  Without the dissipation, the eigenfunctions cross at the EPs.  The substrate dissipation breaks the $\cal PT$ symmetry.  Same parameters as those in Fig.~\ref{Fig10GLrec2Tdep1}.}
\label{Fig13GLrec2Tdep4}
\end{figure}
The corresponding eigenfunctions are shown in Fig.~\ref{Fig11GLrec2Tdep2}.  The two eigenstates coalesce at the EPs ($d=1.951\,\mu$m and $2.034\,\mu$m).  There are two exceptional points and one diabolic point ($d=1.98\,\mu$m) when the thickness varies from $1.9\,\mu$m to $2.1\,\mu$m.  The characteristics of the eigenvalues in the vicinity of the three bifurcation points are demonstrated in Fig.~\ref{Fig12GLrec2Tdep3}, along with the impact of the substrate dissipation (lower right panel of Fig.~\ref{Fig12GLrec2Tdep3}).  An avoided level crossing occurs at the point $d=1.98\,\mu$m (upper right panel).  The phase transitions are associated with the ${\cal PT}$ symmetry of the system.  Adding the substrate dissipation, the system no longer satisfies the ${\cal PT}$ symmetry, and thus the bifurcation points and associated degeneracy disappear as shown in the lower right panel of Fig.~\ref{Fig12GLrec2Tdep3}, but the pure loss (blue curve) and gain (red curve) eigenmodes still exit. 
Figure~\ref{Fig13GLrec2Tdep4} illustrates the eigenfunctions in the complex plane with and without the metallic loss.  When the system undergoes the phase transitions, the eigenfunctions loop around in the complex plane and across at the EPs.  Adding the dissipation breaks the loops.  The left- and right-hand eigenfunctions form a mirror image with respect to the imaginary axis.  Without the background loss, the system satisfies the ${\cal PT}$ symmetry, the background dissipation breaks the anti-symmetric distribution of the imaginary part of the permittivity and lifts the degeneracy and coalescence of the eigenvalues and eigenstates.

\section{Lasing ${\cal PT}$-Spaser}
Although the two-mode interference induces a series of symmetry-breaking transitions, the net amplification does not occur in the output signals.  The sum of transmittance and reflectance remains one as long as the metallic loss is zero.  The ${\cal PT}$-synthetic plasmonic metafilm is in fact a type of spaser\cite{Stockman}.  Without the substrate dissipation, the photons are trapped in the dark modes.  The metallic dissipation provides a mean to couple the ${\cal PT}$-Spaser to free space radiation.

Subwavelength hole array in the metallic substrate supports localized surface plasmon resonances.  The metafilm provides a array of resonators for the localized surface plasmon (SP) of which the electric field parallel to the short edge of the rectangular holes can be excited inside the cavities.  The ${\cal PT}$-synthetic arrangement sustains the gain-SP at the inner walls between the metal and the gain elements and the loss-SP at the inner walls between the metal and lossy elements.  The gain material inside the cavities provides stimulate emission for the gain-SP which compensates its counter-part, the loss-SP.  The dark SP modes having high field strength can store energy and eventually the gain will be saturated.  Even with the gain saturation, the local fields are still much higher than those in conventional plasmonic resonance\cite{Lawandy}.  

Each unit cell in the metafilm contains a pair of balanced oscillating gain-SP and loss-SP.  The coupled ${\cal PT}$-SP-dimer as a whole acts as a meta-atom.  But, unless conventional meta-atoms, the ${\cal PT}$-SP-dimer is actually a {\it gain} meta-atom when the critical level of the gain-loss coupling is exceeded.  Simulation reveals that this critical value is 0.015 for our case.  Above this value, the ${\cal PT}$-metafilm itself can function as a meta-gain medium with the meta-gain atoms made from the ${\cal PT}$-synthetic plasmonic dimers (${\cal PT}$-SP-dimers).  Moreover, the effective gain of this new meta-gain material can be much higher than the original gain material due to pronounced local field enhancement from the dark modes of the ${\cal PT}$-SP-dimers.  

Adding the substrate dissipation is equivalent to introducing loss into the cavity.  The dissipation provides a coupling mechanism for the spaser to radiate into the far field.  Without the dissipation, the ${\cal PT}$-Spaser will not lase and behaves as a lossless thin film.  The Spaser generates strong coherent local field, but the photons are trapped inside the dark modes which cannot radiate without a proper coupling mechanism.  Note that here the loss is introduced into the cavity as metallic dissipation which naturally enters the plasmonic dark modes and turns them into leaky modes, and thus releases the trapped photons.  

The top and bottom surfaces of the metafilm provide the required feedback for lasing.  This feedback mechanism is observable from Fig.~\ref{Fig2GLrec2Tz3} that shows the Fabry P\'{e}rot effect.  The vanishing $M_{11}$, a necessary condition for the super scattering to occur, is in fact the condition of lasing threshold\cite{Longhi3}.  The spectral singularity in transmission and reflection is inherently associated with a laser.  Once the photons are released from the dark modes, the identical ${\cal PT}$-dimers in the planar periodic array oscillate in phase due to mutual coupling and lead to spatially and temporally coherent emission normal to the film (analogy to phased antenna array).  The metafilm as a whole functions as a high quality active resonator.
The lasing mechanism described here is different from those reported recently\cite{Plum,Beijnum}.  In their experiments, the gain medium and plasmonic material are two separate media and the out-coupling method is also different.  Here, the metafilm itself acts as a meta-gain medium with the meta-gain atoms made from the $\cal PT$-SP-dimers.  The $\cal PT$-Spaser is coupled to the far field through transforming dark modes into leaky modes by metallic dissipation.  The plasmonic $\cal PT$-laser revealed here is also different from those $\cal PT$-waveguide lasers\cite{Feng3,Hodaei} which select single mode operation based on $\cal PT$ symmetry breaking.  Here, the lasing ${\cal PT}$-Spaser can potentially generate much stronger radiation with spatial and temporal coherence.

\section{Summary}
We have demonstrated a new type of plasmonic laser based on the $\cal PT$ symmetry, lasing $\cal PT$-spaser.  The super scattering has a finite bandwidth that increases the detection possibility for spectral singularity related resonances.  Our result implies that the scattering properties of optical systems may be controlled by the background dissipation.  Our result may provide a new strategy to control the phase of the $\cal PT$ symmetry using the background loss, and to control the super radiation from a cavity by tuning the cavity dissipation.  The effect of introducing gain into materials is more than loss compensation.  The interplay between the loss and gain SPs in the dimers leads to the meta-gain atoms ($\cal PT$-SP-dimers) which transform the $\cal PT$-synthetic metafilm into a meta-gain medium.  Our numerical experiment is one step further toward the practical $\cal PT$-synthetic materials.  The $\pi$-phase jump of the electric field at the super scattering can be used to design a two-state system, a new modulation scheme, and potentially manipulating light in free space using the concept of $\cal PT$ symmetry.  
The narrow-band resonant thinfilms may find applications in the areas of notch filter, sensor, and optical switch.

\acknowledgements
%The author acknowledges the support from K. Boulais and gratefully thanks the referee for his/her invaluable comments.  This project is funded by Office of Naval Research and In-House Applied Research program at NSWC, Dahlgren.
The author acknowledges the support from K. Boulais.  This project is funded by Office of Naval Research and In-House Applied Research program at NSWC, Dahlgren.

%NSWCDD-PN-15-00056 is approved for Distribution Statement A: Approved for Public Release; distribution is unlimited.

\end{document}